\documentclass[12pt,draftcls,onecolumn]{IEEEtran}
\usepackage{hyperref}
\usepackage{amsmath,amssymb}
\usepackage{latexsym}
\usepackage{CJK}
\usepackage{cite}   %ÎÄÏ×ÒýÓÃ
\usepackage{caption} %ͼƬÎÄ×Ö¼ä¾à

\usepackage{verbatim}  % \begin{comment} ... \end{comment} °üº¬±»×¢Ê͵ÄÎÄ×Ö
\usepackage{graphicx}  % ²åÈëͼƬ
\usepackage{bm}  %Êýѧ»·¾³Ï¼ӴÖ×ÖÌå$\bm{ a }$
\usepackage{mathrsfs} % »¨Ìå×Öĸ
\usepackage{algorithm} %format of the algorithm
\usepackage{algorithmic} %format of the algorithm
\usepackage{booktabs}
\usepackage{textcomp}  %ǧ·ÖºÅ\textperthousand ÉãÊ϶ȣº$^\circ$C \par
\usepackage{multirow}  %¶àÐкϲ¢
\usepackage{lettrine}   %Ê××ÖϳÁ
\usepackage{graphicx}  %Ë«À¸Ä£°å²åÈëͨÀ¸Í¼Æ¬
\usepackage{color}  %×ÖÌåÑÕÉ«
\usepackage{amsmath}
\usepackage{amssymb}

\DeclareMathOperator*{\argmin}{argmin}

\usepackage{algorithm}
\usepackage{algorithmic}

\usepackage{amsthm} %thm / proof»·¾³µÈÊýѧ»·¾³

\theoremstyle{plain}

\newtheorem{lem}{Lemma}
\newtheorem{prop}{Proposition}

\theoremstyle{definition}
\newtheorem{defn}{Definition}

\theoremstyle{remark}
\newtheorem*{rem}{Remark}

\newcommand{\myvec}[1]%
   {\stackrel{\raisebox{-2pt}[0pt][0pt]{\small$\rightharpoonup$}}{#1}}

\begin{document}
%
% paper title
% can use linebreaks \\ within to get better formatting as desired
\title{Directivity-Beamwidth Tradeoff of Massive MIMO Uplink Beamforming for High Speed Train Communication}

% author names and affiliations
% use a multiple column layout for up to two different
% affiliations

\author{Xuhong Chen$^1$, Jiaxun Lu$^1$, Tao Li$^1$, Pingyi~Fan$^1$ and~Khaled~Ben~Letaief$^2$~\IEEEmembership{Fellow,~IEEE}\\
        $^1$State Key Laboratory on Microwave and Digital Communications, \\
        Tsinghua National Laboratory for Information Science and Technology, \\
        Department of Electronic Engineering, Tsinghua University, Beijing, China.\\
        $^2$Department of Electrical and Computer Engineering, Hong Kong University of Science and Technology, Hong Kong.\\
\thanks{Corresponding Author is Pingyi Fan.}
E-mail: chenxh13@mails.tsinghua.edu.cn, lujx14@mails.tsinghua.edu.cn, \\
        litao12@mails.tsinghua.edu.cn, fpy@mail.tsinghua.edu.cn, eekhaled@ust.hk
}

\maketitle

\begin{abstract}
High-mobility adaption and massive Multiple-input Multiple-output (MIMO) application are two primary evolving objectives for the next generation high speed train (HST) wireless communication system. In this paper, we consider how to design a location-aware beamforming for the massive MIMO system in the high traffic density HST network. We first analyze the tradeoff between beam directivity and beamwidth, based on which we present the sensitivity analysis of positioning accuracy. Then, in order to guarantee a high efficient transmission, we derive an optimal problem to maximize the beam directivity under the restriction of diverse positioning accuracies. After that, we present a low-complexity beamforming design by utilizing location information, which requires neither eigen-decomposing (ED) the uplink channel covariance matrix (CCM) nor ED the downlink CCM (DCCM). Finally, we study the beamforming scheme in future high traffic density HST network, where a two HSTs encountering scenario is emphasized. By utilizing the real-time location information, we propose an optimal adaptive beamforming scheme to maximize the achievable rate region under limited channel source constraint. Numerical simulation indicates that a massive MIMO system with less than a certain positioning error can guarantee a required performance with satisfying transmission efficiency in the high traffic density HST scenario and the achievable rate region when two HSTs encounter is greatly improved as well.
\end{abstract}


\IEEEpeerreviewmaketitle

\section{Introduction}
\begin{figure}[!b]
\centering
\includegraphics[width=0.45\textwidth]{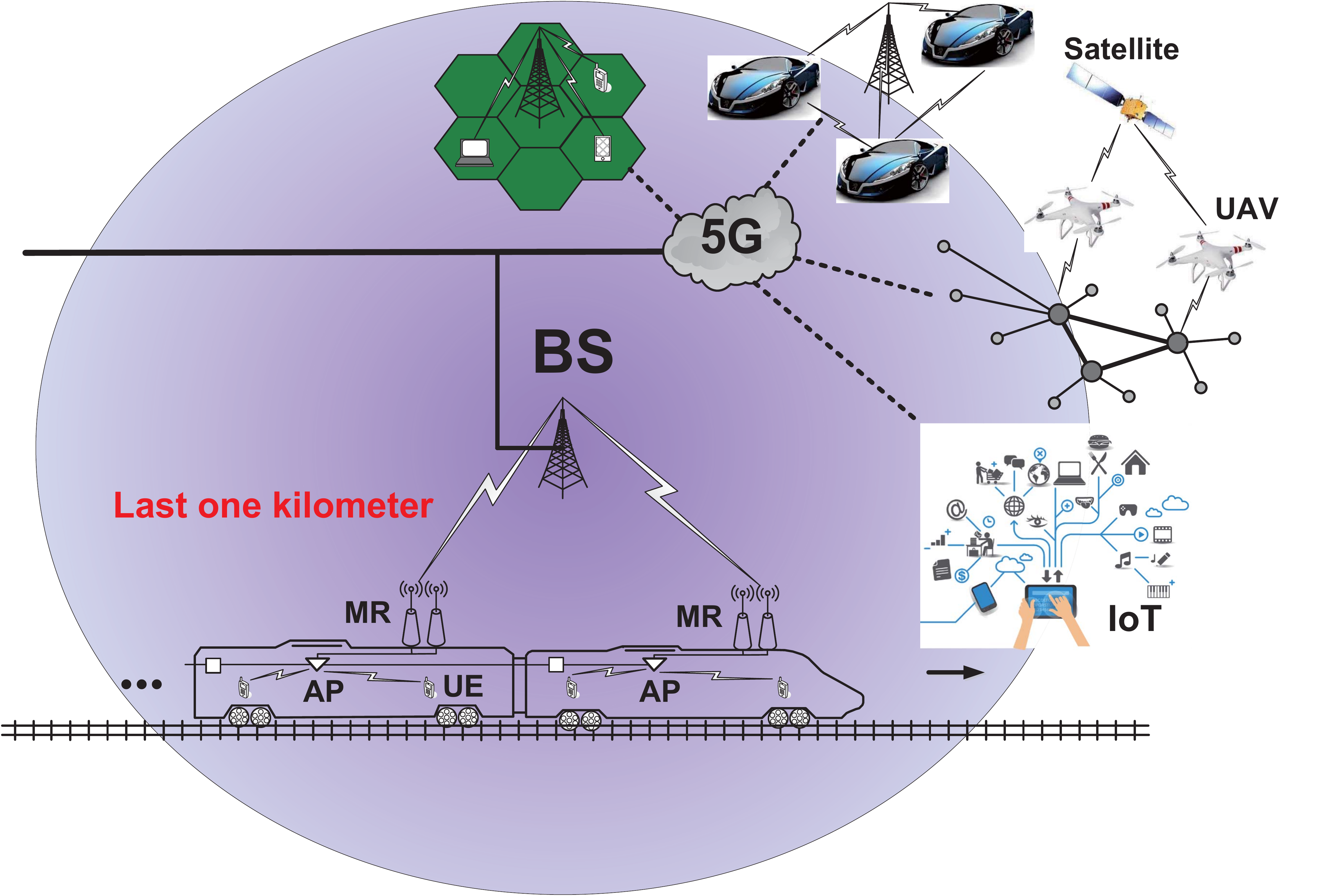}
\caption{The integration of 5G network and the HST wireless communication system.} \label{fig:5G}
\captionsetup{belowskip=-10pt}
\end{figure}

% no \IEEEPARstart
\lettrine[lines=2]{H}igh speed train (HST) wireless communication is an important component in future 5G wireless communication networks \cite{5G} because high mobility adaption is one of the key evolution objectives for 5G. Therefore, as illustrated in Fig. \ref{fig:5G}, the integration of future 5G cellular networks and the HST wireless communication network is paramount to provide a fast and seamless wireless service for users on the HST, where the last one kilometer communication between the HST and the wayside base station (BS) plays a key role to guarantee the quality of service (QoS) for users onboard \cite{seamless}. To this end, new advanced physical layer access technologies need to be invented and employed for the future high traffic density HST network according to the trend of 5G system and the character of HST scenario.

In the future 5G system, massive Multiple-input Multiple-output (MIMO) is deemed as a prominent technology to improve the spectrum efficiency \cite{5G1} through exploiting the multiplexing and diversity gain, where the appropriately-designed beamforming is utilized to diminish interference \cite{beamform}. Previous works in \cite{work2, work3, work4} employing diverse beamforming schemes in high mobility scenario demonstrated a significant performance improvement by utilizing the directional radiation. However, neither of them considered how to reduce the implementation complexity of fast beamforming for massive MIMO system in HST scenario, especially when the instant channel state information (CSI) can not be obtained accurately. That is, the large online computational complexity is hard to be carried out in the process of channel covariance matrix (CCM) acquisition and eigen-decomposing (ED) CCM when adopting conventional massive MIMO beamforming.

As to the wireless communication system supporting future high traffic density HST network, high mobility adaption is one of the crucial evolution targets \cite{mobility, mobility1}, where the wireless transmission demand is ever-growing according to the real-scenario estimation in \cite{demand} (the estimated demands could be as high as 65Mbps over a bandwidth of 10MHz for a train with 16 carriages and 1000 seats). In the literature, certain diverse designs aiming to improve the user QoS in HST scenario have been proposed \cite{work, work1}. However, low-complexity beamforming designs for high mobility scenarios are still under-developed. Besides, to the best of our knowledge, previous studies mainly focus on the low traffic density scenario. Particularly, as a typical example, a single HST scenario is widely studied, where the wayside BS merely serves one HST for the whole HST traversing period. But, in future high traffic HST network, a two HSTs encountering scenario is ubiquitous, where the two HSTs have to share the limited wireless channel resources and a performance deterioration can be anticipated without appropriate beam allocation strategies. Thus, the low-complexity beamforming scheme should be designed according to the specific character of the HST scenario.

\begin{figure}[!t]
\centering
\includegraphics[width=0.45\textwidth]{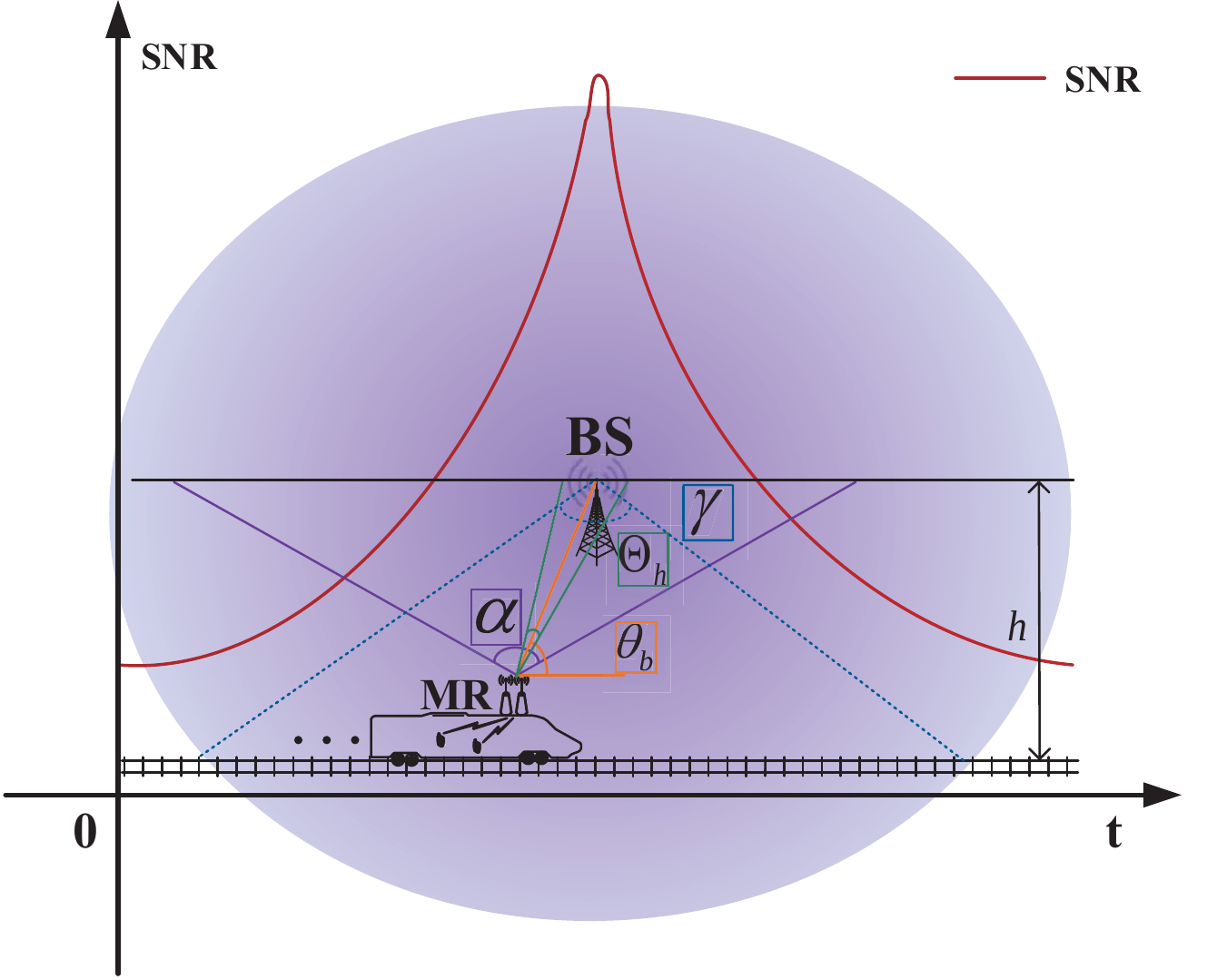}
\caption{The HST scenario and beam coverage.} \label{fig:hsr}
\captionsetup{belowskip=-10pt}
\end{figure}
In the HST scenario, the special scenario characters distinguish this scenario from conventional low mobility scenario. As illustrated in Fig. \ref{fig:hsr}, when the HST quickly traverses the coverage area of one BS, the received signal-to-noise ratio (SNR) at the mobile relay (MR) will fluctuate dramatically due to the large variation of path loss, which makes the received SNR in the edge area of the BS coverage quite low, bringing challenges for conventional adaptive beamforming \cite{abf} or orthogonal switched beamforming \cite{obf} in channel detections. The estimated maximum Doppler shift with carrier frequency at 2.35 GHz will be 945 Hz when the train velocity is 486 km/h \cite{demand}, which implies that the channel coherence time is less than 1 ms. Consequently, it is hardly to track the channel in this scenario. Moreover, the complex channel environment due to traversing diverse terrains (typical scenarios like viaduct, mountain, etc.) makes it difficult to accurately estimate the channel with low cost since the wireless channel appears fast time-varying and double-selective fading in spatial-temporal domains. On the other hand, even if the CSI is acquired, conventional beamforming designs for massive MIMO system in low mobility scenario can not be directly applied here since it will lead to undesirable performance. To this end, fast beamforming scheme for massive MIMO system in this scenario requires to be redesigned with a less complicated method.

Although high mobility causes new challenges in the application of beamforming scheme, if we take advantage of high mobility in a different perspective, the drawback can be transformed into valuable side information. Namely, thanks to the safety guarantee for HST, the location of the HST can be trackable and predictable because the moving trend will not change in a short time. Actually for the HST scenario, the train can only move along the pre-constructed rail rather than random movement, which indicates no spatial-random burst communication requests will occur. That is, the entrance time of the HST is predictable and the wireless communication requests only comes from one side of the BS coverage, which narrows down the coverage scope of the beamforming scheme.

In this paper, we introduce a simple low-complexity beamforming scheme for massive MIMO system in HST scenario by exploiting the HST location information. Because the conduction of this beamforming scheme required neither ED uplink CCM (UCCM) nor downlink CCM (DCCM) and therefore, the aforementioned challenges in channel detections and large online computational complexity can be alleviated. Since the HST location information plays a crucial role in the beamforming scheme, we first analyze the tradeoff between beamwidth and directivity in high mobility scenario and find that it is independent of antenna spacing and total beam number. Then, we present the sensitivity analysis of positioning accuracy against beam directivity and formulate it as an optimization problem. Then, we also present an optimal algorithm to slove it. Finally, taking a two HSTs encountering scenario as a typical example of the HST network, we improve the fast beamforming scheme in terms of achievable rate region with the help of HST location information.

The contributions of this paper can be concluded as

1)A tradeoff analysis between beam directivity and beamwidth of the fast beamforming scheme for HST scenario is presented.

2)A location-sensitivity analysis and an optimal beamforming solution to maximize the directivity for a given error range or error distribution of the location information are provided.

3)A location-aware low-complexity beamforming scheme for the HST massive MIMO system is given.

4)An explicit closed-form expression of the optimal resource allocation for uplink beamforming to maximize the achievable rate region in high traffic density HST network is presented.
% \begin{itemize}
%\item \emph{\textbf{A tradeoff analysis between beam directivity and beamwidth of the fast beamforming scheme for HST scenario is presented.}}
%\item \emph{\textbf{A location-sensitivity analysis and an optimal beamforming solution to maximize the directivity for a given error range or error distribution of the location information are provided.}}
%\item \emph{\textbf{A location-aware low-complexity beamforming scheme for the HST massive MIMO system is given.}}
%\item \emph{\textbf{An explicit closed-form expression of the optimal resource allocation for uplink beamforming to maximize the achievable rate region in high traffic density HST network is presented.}}
%\end{itemize}

The rest of this paper is organized as follows. Section II introduces the system model and the transceiver structure. Section III presents the directivity-beamwidth tradeoff analysis and the optimization solution to maximize the directivity under diverse positioning error constrains. Section IV presents the application process of designed low-complexity beamforming scheme for HST scenario. In Section V, we exhibits the location-aided adaptive beamforming strategy for the high traffic density HST network. Section VI shows the numerical simulation results and the corresponding analyses. Finally, we conclude it in Section VII.

\section{System Model}

\begin{figure}[!t]
\centering
\includegraphics[width=0.5\textwidth]{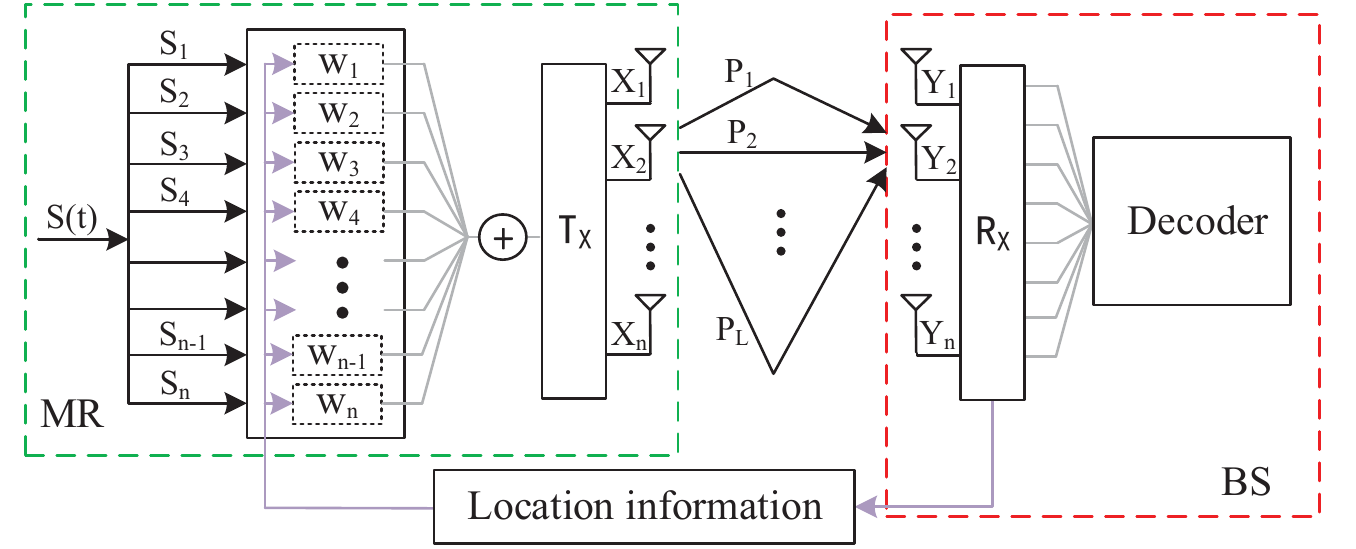}
\caption{The transceiver structure for HST massive MIMO systems.} \label{fig:transceiver}
\captionsetup{belowskip=-10pt}
\end{figure}

Let us consider the uplink beamforming transceiver structure between one carriage and the BS, depicted in Fig. \ref{fig:transceiver}, where we omit most parts of baseband transformations and only focus on the beamforming part. The user equipment (UE) inside the train connects to the access point (AP) inside the carriage and the information packet will be forwarded to wayside BS by the MR mounted on the top of the HST to avoid large penetration loss. The MR is equipped with $M$-element uniform linear single-array antenna due to the \emph{3dB} gain over double-array structure \cite{gain}. To the link between BS and the MR, the Doppler effect in HST scenario will not be considered here since it can be accurately estimated and removed from the signal transmission part as shown in \cite{work1}. Let the antenna spacing and carrier wavelength be $d$ and $\lambda$, respectively. For a multi-element uniform linear array, where $d \gg \lambda$, the half-power beamwidth $\Theta_h$ (in Fig. \ref{fig:hsr}) can be equivalently expressed as \cite{antenna}

\begin{equation}\label{F1}
\Theta_h = \frac{C \lambda}{\pi d N},
\end{equation}
where $C=2.782$ represents a constant parameter in antenna design and $N$ ($N\le M$) is the total generated beam number.

In HST scenario, the viaduct scenario occupies 80 percent of the entire route \cite{demand}, which makes the line-of-sight (LOS) signal dominant and the angular spread around the MR relatively tight. In addition, the instantaneous location information $\theta_b$ (shown in Fig. \ref{fig:hsr}) of BS can be acquired by Global Positioning Systems, accelerometer and monitoring sensors along the railway or, can be estimated according to the entrance time as \cite{work4}. Thus, a location-aware beamforming can be carried out by exploiting $\theta_b$. In this way, it can reduce the beamforming complexity, which will be presented hereinafter.

As shown in Fig. \ref{fig:hsr}, the total beamwidth, which is a constant for a deployed BS, can be expressed by
\begin{equation*}
 \alpha=N \Theta_h = \frac{C \lambda}{\pi d}.
\end{equation*}

To enhance the coverage of the MR on the train, $\alpha > \delta$ is essential, where $\alpha$ represents the beam coverage angle of the HST and $\delta$ is the coverage angle of BS. During the traversing period of the HST, the generated beam will cover the location of the BS with appropriate beamwidth to guarantee the uplink transmission.
%The simplified transceiver structure between one carriage and the BS is illustrated in Fig. \ref{fig:transceiver},  The beam weight is pre-calculated and pre-stored according to different location along the rail, which will be explained later in Section V. The received signal strength $P_r$ can be expressed as
%\begin{equation}\label{Pr}
%P_r=\sum\limits_{n=1}^{N_s}\overline{P}G(t)w^2_nh(t),
%\end{equation}
%where $\overline{P}$ is the total power at time $t$ and $w_n$ is the selected beam weight according to location information $\theta_b$. $G(t)$ and $h(t)$ respectively denote the large-scale and small-scale fading. $N_s$ represents the total selected beam number.

\section{Directivity-Beamwidth Tradeoff and Efficient Transmission}\label{Sec:BeamAndConnection}

Theoretically, adding more antennas (elements) can improve the beamforming performance due to the increased directivity. However, it also narrows down the beam width. In practice, positioning error may occur due to some reasons, which may degrade the performance of the location-aware low-complexity beamforming scheme and trigger low transmission efficiency. In this section, we first analyze the tradeoff between beam directivity and beamwidth and then, derive the efficient beamforming probability under diverse positioning error constrains, where the effective beamforming is defined as the BS is in the coverage of the selected beam. Finally, under the precondition of efficient transmission, we maximize the beam directivity with error-constrained location information.

The corresponding directivity of each beam in massive MIMO system, i.e. $N\pi d/\lambda$ is sufficiently large, can be expressed as

\begin{equation}\label{F2}
D = \frac{TdN}{\lambda},
\end{equation}
where $T$ depends on the specific type of linear array (i.e. $T=2$ for broadside array and $T=4$ for ordinary end-fire array).

Thus, the relations between beam directivity and beamwidth can be expressed in the following lemma.

\begin{lem}For a given linear antenna array, its designed beam directivity and bandwidth can be adjusted, then the relationship between them are give as follows.

\begin{equation}\label{F3}
D = \frac{TC}{\pi \Theta_h}.
\end{equation}

\end{lem}

\begin{rem}
The tradeoff between beam directivity and beamwidth is independent of the antenna spacing $d$ and the beam number $N$. That is, the variation of $d$ and $N$ only affects the value of beam directivity and beamwidth, but has no influence on the ratio of the beam directivity and beamwidth.
\end{rem}
Observing Eq. (\ref{F1}) -(\ref{F3}), one can find that $d$ and $N$ are dual with respect to $\Theta_h$ and $D$. That is, the variation of $d$ $\rightarrow$ $d^{'}$ is equivalent to $N$ $\rightarrow$ $N^{'}$, following the relationship
$$
\frac{d}{d^{'}} = \frac{N^{'}}{N}.
$$

As shown in Fig. \ref{fig:hsr}, let the BS's relative location and the vertical distance between BS and rail be denoted by $\theta_b$ and $d_0$, respectively. If the BS is in the $i$-th beam, where

\begin{equation}\label{equ:index}
\begin{split}
i &= \biggl\lfloor\frac{2\theta_b + \alpha - \pi}{2\Theta_h}\biggr\rfloor\\
& = \biggl\lfloor\frac{2\theta_b - \pi}{2\Theta_h}\biggr\rfloor + \frac{N}{2}.
\end{split}
\end{equation}

\begin{figure}[b]
\centering
\includegraphics[width=0.48\textwidth]{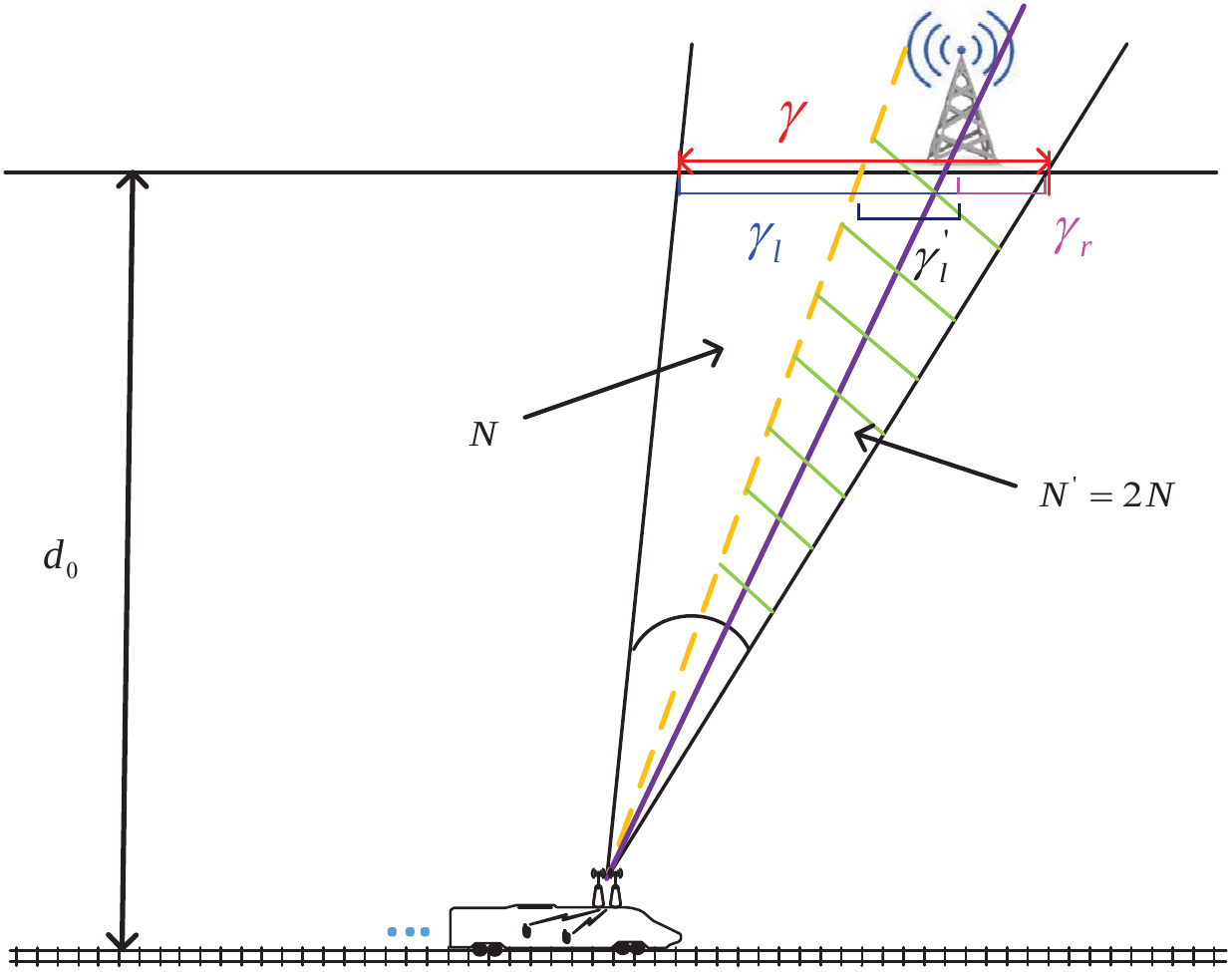}
\caption{The beam devision process.} \label{Fig:BeamDevide}
\end{figure}

Then, as illustrated in Fig. \ref{Fig:BeamDevide}, the distances between left(right) bounds (denoted as $\gamma_l$($\gamma_r$)) and the BS of the $i$-th beam can be expressed by

\begin{equation}\label{equ:gamma_l}
\gamma_l = \frac{[\pi+2\chi\Theta_h-2\theta_b]d_0}{2\sin \theta_b}%\frac{[\pi+(2-2\chi)\Theta_h - 2\theta_b]h}{2 \sin \theta_b}
\end{equation}
and

\begin{equation}\label{equ:gamma_r}
\gamma_r = \frac{[2 \theta_b-\pi-2(\chi-1)\Theta_h]d_0}{2 \sin \theta_b},%\frac{[\pi - 2\chi\Theta_h - 2\theta_b]h}{2 \sin \theta_b},
\end{equation}
respectively, where $\chi = \biggl\lfloor\frac{2\theta_b - \pi}{2\Theta_h}\biggr\rfloor$.

Therefore, the corresponding coverage length on the BS side of the $i$-th beam is given by

\begin{equation}
\gamma = \gamma_l + \gamma_r \approx \frac{d_0}{\sin \theta_b} \Theta_h = \frac{C \lambda d_0}{\pi d N \sin \theta_b}.
\end{equation}

The approximation shown in Eq. (7) is because our considered $\Theta_h$ is usually small to ensure high channel gain in one beam. Now let us consider the case there exists positioning error $\Delta x$, which may lead to a deterioration of the effective beamforming performance. In fact, $\Delta x$ may be caused by GPS estimation errors, quantization errors, random scattering of electronic waves, etc. For simplicity, it is assumed that $\Delta x$ is Gaussian distributed in the sequel, i.e. $\Delta x \sim \emph{N}(0,\sigma^2)$. We shall utilize it in the following analysis and optimization.

With the widely used Q function in signal detection, i.e. $Q(x) = \int_{x}^{+\infty} \frac{1}{\sqrt{2 \pi}} exp(-\frac{t^2}{2})dt$, the effective beamforming probability $P_i$ can be expressed by

\begin{equation}
P_i = 1 - \frac{Q(\frac{\gamma_l}{\sigma}) + Q(\frac{\gamma_r}{\sigma})}{2}.
\end{equation}

For a given effective beamforming probability threshold $P_{th}$, the beamforming design problem can be formulated as the following optimization problem

\begin{align*}\label{equ:formulation}
\tag{9}&\max \,\, D\\
\tag{10}&s.t.\quad  P_i \geq P_{th},
\end{align*}
where we try to maximize the directivity (beam gain) under the constraint condition (10). To solve the maximization problem above, we first analyze the monotonicity of $P_i$. As shown in Fig. \ref{Fig:BeamDevide}, the original total generated beam number is $N$ with the coverage length $\gamma=\gamma_l+\gamma_r$. If the location information is accurate enough, a beam with high directivity can be generated by increasing the total generated beam number from $N$ to $N^{'} = 2N$. Then, the coverage length for the new beam is $\gamma_l^{'}+\gamma_r$. After the beam division process, $\gamma_r$ is unchanged, then $\gamma_l$ has been shorten to $\gamma_l^{'} = \gamma_l - \frac{\gamma}{2}$ and $P_i^{'} = P_i - \frac{Q(\gamma_l^{'}) - Q(\gamma_l)}{2}$. That indicates the $P_i$ monotonous decreases with respect to $N$. Therefore, the solving of (9) can be turned into a searching problem as follows

\begin{equation}\label{equ:minError}
N^*=\mathop{\argmin}_N (P_i - P_{th}) \tag{11}.
\end{equation}

Note that the directivity $D$ increases in proportion to $N$.

In literature, there are many techniques to solve (11). Considering the fact that number of available $N$ is finite in a system, the searching process can be summarized as follows.

\begin{algorithm}\label{alg:search}
\caption{The searching method of $N^*$}
\begin{algorithmic}
\STATE {set $N^*=1$}
\REPEAT
\STATE 1. generate i, $\gamma_l$ and $\gamma_r$ with \eqref{equ:index}, \eqref{equ:gamma_l} and \eqref{equ:gamma_r}.
\STATE 2. $1 - \frac{Q(\gamma_l) + Q(\gamma_r)}{2} \rightarrow P_i$.
\STATE 3. if $P_i > P_{th}$, $N^* = 2N^*$.
\UNTIL{$P_i \leq P_{th}$}
\STATE \textbf{OUTPUT} $N^*$.
\end{algorithmic}
\end{algorithm}

It can be noted that, when the train location angle $\theta_b = \frac{\pi}{2}$, the adjusting of $N$ would not improve effective beamforming probability significantly. In that case, $\gamma_l$ or $\gamma_r$ is limited by system structure instead of beam number $N$, which needs further considerations. That is, the optimal beam number $N^*$ searching method can be extended to scenarios with $\theta_b \in (\frac{\pi-\alpha}{2}, \frac{\pi}{2})\cup (\frac{\pi}{2}, \alpha)$.

\section{Application of Location-aware Low-complexity Uplink Beamforming for Massive MIMO System}

\subsection{The Beam Generation Process}
Directional beam can be generated through diverse phase excitations on each element according to \cite{antenna, directional}. Let the total generated beam number be $N$ (depends on the BS deployments),    the beam weight of the $i$-th beam ($i=1,2,...,N$) is defined as

\begin{equation}\label{weight}
w_i=f_iD_i(\theta_{b})\tag{12},
\end{equation}
where $f_i=\sum\nolimits_{m=1}^Mf_i(m)$ denotes the power allocation coefficient for the $i$th beam and $f_i(m)$ stands for the actual amplitude excitation on the $m$-th ($m=1,2,...,M$) element for the $i$-th beam. Usually, for a uniform linear array the amplitude excitation is equal on each element. $D_i(\theta_{b})$ is the directivity of the selected $i$-th beam according to the location information $\theta_{b}$.

If we label the phase excitation for $i$-th beam as $\bm{\beta}^m_i$, the corresponding uplink steering vector on each element for an acquired location information $\theta_{b}$ can be denoted as

\begin{equation*}
\bm{e}_i(\theta_{b})=\left [1,e^{j(kd\cdot \cos\theta_{b}+\beta_i^2)}, \ ... \ ,e^{j(M-1)(kd\cdot \cos\theta_{b}+\beta_i^M)} \right ]^T,
\end{equation*}
where $k=\frac{2\pi}{\lambda}$.

\begin{figure}[!t]
\centering
\includegraphics[width=0.5\textwidth]{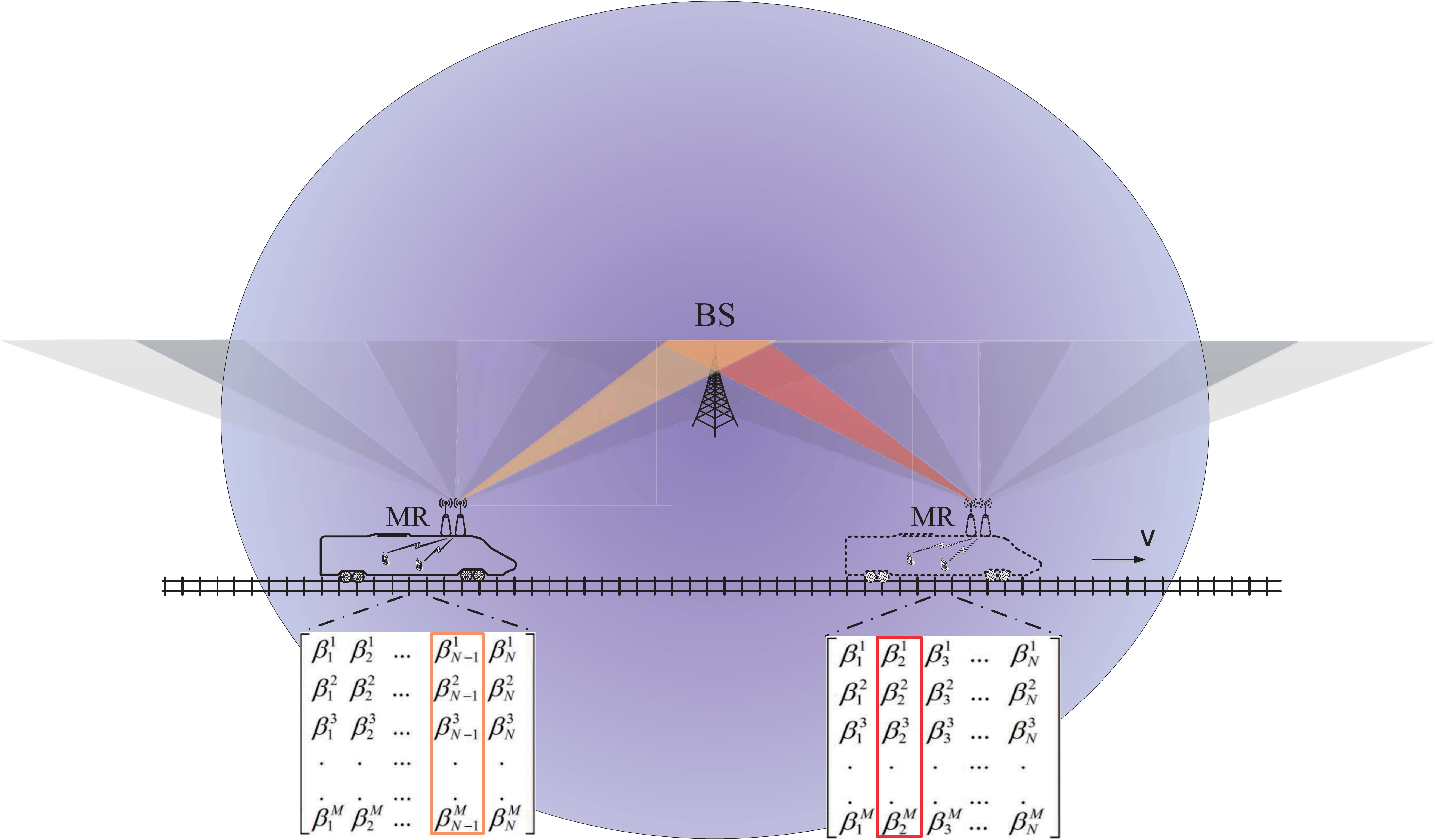}
\caption{The location-aware beamforming process.} \label{fig:process}
\captionsetup{belowskip=-10pt}
\end{figure}

\begin{rem}
For deployed antenna array, one can off-line calculate and pre-set the phase excitation at the MR according to the different but restricted location information $\theta_{b}$, $\theta_{b}\in \left [\frac{\pi}{2}-\frac{\alpha}{2}, \frac{\pi}{2}+\frac{\alpha}{2}\right ]$. The pre-calculated phase excitation for $i$-th beam is

\begin{equation}\label{phase1}
\bm{\beta}_i=\left [\beta_i^1, \beta_i^2, \ ... \ , \beta_i^M \right ]^T\tag{13},
\end{equation}
and the whole mapper for each beam directed to diverse locations on the rail can be expressed as
\begin{equation}       %¿ªÊ¼Êýѧ»·¾³
\bm{\beta}=\left[                 %×óÀ¨ºÅ
  \begin{array}{ccccc}   %¸Ã¾ØÕóÒ»¹²5ÁУ¬Ã¿Ò»Áж¼¾ÓÖзÅÖÃ
    \beta_1^1 & \beta_2^1 & \beta_3^1 & ...  & \beta_N^1\\  %µÚÒ»ÐÐÔªËØ
    \beta_1^2 & \beta_2^2 & \beta_3^2 & ...  & \beta_N^2\\  %µÚ¶þÐÐÔªËØ
    \beta_1^3 & \beta_2^3 & \beta_3^3 & ...  & \beta_N^3\\
      \cdot   &   \cdot   &   \cdot   & ...  &   \cdot  \\
    \beta_1^M & \beta_2^M & \beta_3^M & ...  & \beta_N^M\\
  \end{array}\tag{14}
\right].                 %ÓÒÀ¨ºÅ
\end{equation}
\end{rem}

\begin{algorithm}[!t]\label{alg:Proc}
\caption{The location-aware beam selection procedure}
\begin{algorithmic}[1]
\label{alg:Proc}
\STATE \textbf{INPUT} $\theta_{b}$.
\STATE {Initialize $\bm{\beta}_c=[\ ]$.}
\REPEAT
\STATE 1. generate N with \textbf{Algorithm 1}.
\STATE 2. acquiring $\theta_{b}$.
\STATE 3. if $\theta_{b}\in \left[ \left(\frac{\pi}{2}-\frac{\alpha}{2}\right)+(i-1)\frac{\alpha}{N}, \left (\frac{\pi}{2}-\frac{\alpha}{2}\right )+i\frac{\alpha}{N}\right ]$, $\bm{\beta}_c=\bm{\beta}_i$.
\UNTIL{$\theta_{b}\ge \frac{\pi}{2}+\frac{\alpha}{2}$}
\STATE {set $\bm{\beta}_c=\bm{\beta}_1$.}
\STATE \textbf{OUTPUT} The phase excitation for current location $\bm{\beta}_c$.
\end{algorithmic}
\end{algorithm}

\subsection{The Beam Selection Process}

In the beam selection process, we only emphasize on the selection between beams that are generated by the same BS and the inter-BS beam selection for handover is not our primary concern. The beam selection is based on the acquired location information, where only one beam can be selected at a time. For example, when the train entrances the coverage of one BS at the first moment, the right-most beam will be selected to transmit data. The location information will be continuously updated and matched with the phase excitation mapper as the train keeps moving along the time. When the train moves into a new location covered by another beam, the corresponding phase excitation vector will be utilized, which is illustrated as in Fig \ref{fig:process}. Therefore, the whole location-aware beam selection process can be implemented repeatedly, similar to a routing process, where the phase excitation mapper $\bm{\beta}$ is the routing table. The detailed selection process is expressed in Algorithm \ref{alg:Proc}, which requires neither CSI detections nor ED CCM and therefore, greatly simplifying the online computational complexity.

\section{Uplink Adaptive Resource Allocation of Beamforming for High Traffic Density HST Network}

For the future development of HST network, the HST operation speed will be raised up and more HSTs will operate simultaneously to meet the growing transportation capacity. Consequently, a two HSTs encountering scenario will occur most frequently within the transmission coverage of one BS. Therefore, to ensure both of the HSTs can enjoy a good QoS during the encountering process, an appropriate power allocation strategy for the two selected beams needs to be subtly designed.

\begin{figure}[b]
\centering
\includegraphics[width=0.5\textwidth]{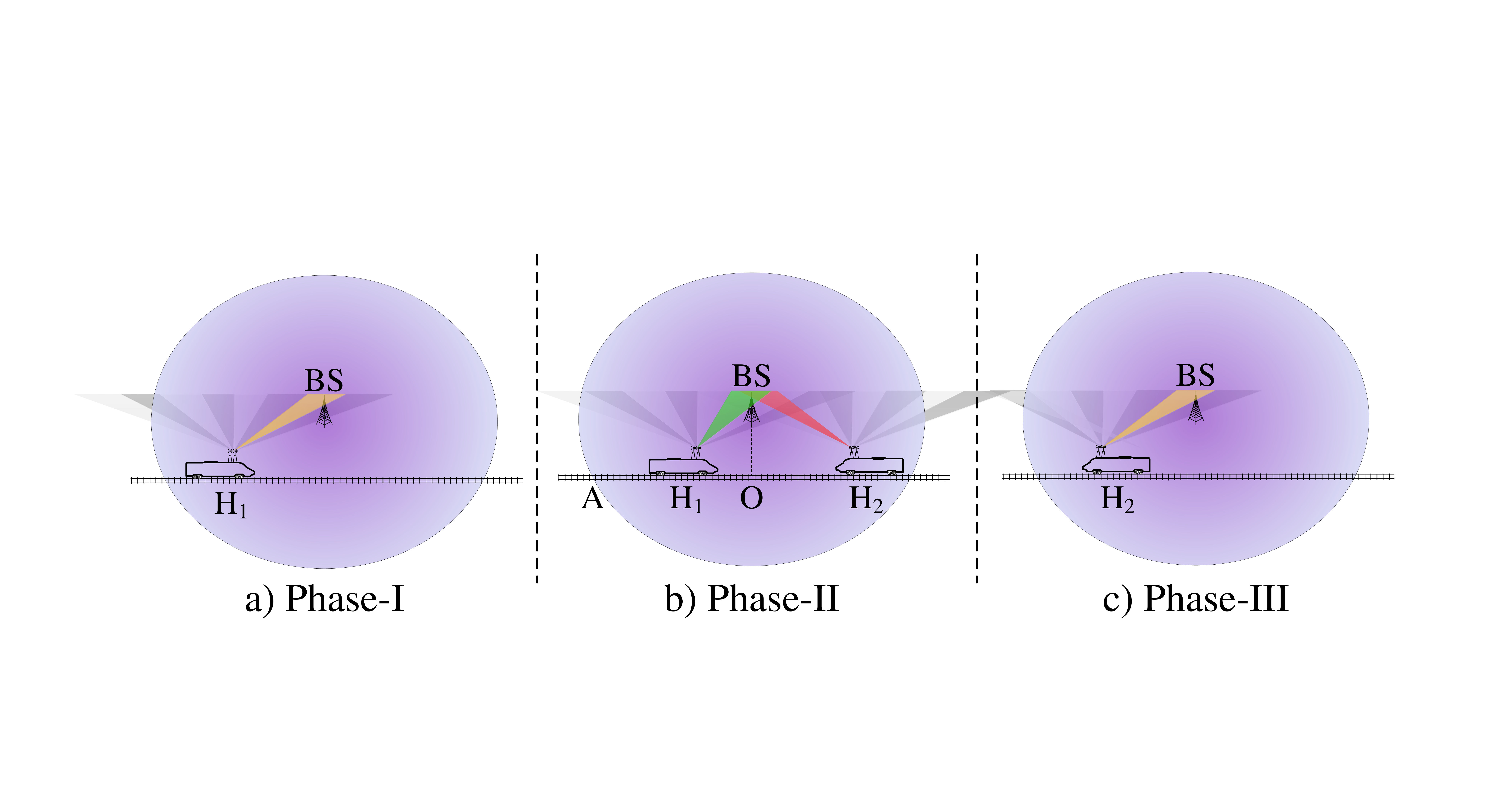}
\caption{The three transmission phases of the encountering process, denoting as $\mathbb{T}_1$, $\mathbb{T}_2$ and $\mathbb{T}_3$.} \label{Fig:Twotrain}
\end{figure}

As illustrated in Fig. \ref{Fig:Twotrain}, two HSTs (denoted as $H_1$ and $H_2$) are running at a constant velocity $v_0$. The performance degradation triggered by encountering process mainly depends on the intersection location of the two HSTs. For example, if the intersection point locates at the center of the rail (point O), two HSTs will compete for the longest time period and thus the deterioration will be most serious due to the channel resource competition. Without loss of generality, we assume that $H_1$ enters the coverage of the BS earlier than $H_2$ and when $H_2$ enters the BS coverage, the distance between $H_1$ and point $A$ is denoted as $\eta L$ ($0\le \eta \le 2$). According to the aforementioned beam selection process, when $H_1$ selects the $i$-th beam at system time $t$, the corresponding beamforming weight is denoted as $w_i^1(t)$ ($w_i^2(t)$ for $H_2$). Based on above assumptions, the encountering process will take place in time $t\in \mathbb{T}=[-\frac{\eta L}{v_0}, \frac{2 L}{v_0}]$, which can be divided into three phases as $\mathbb{T}_1$, $\mathbb{T}_2$ and $\mathbb{T}_3$.

When $t\in \mathbb{T}_1\cup \mathbb{T}_2$, the HST $H_1$ is served by the BS. The corresponding transmission distance between $H_1$ and BS is

\begin{equation}\label{distance}
d_1(t)=\sqrt{d_0^2+h_0^2+(v_0t-L+\eta L)^2}, t\in \mathbb{T}_1\cup \mathbb{T}_2, \tag{15a}
\end{equation}
where $h_0$ the antenna height at the BS. Similarly, the transmission distance between $H_2$ and BS is

\begin{equation}\label{distance}
d_2(t)=\sqrt{d_0^2+h_0^2+(v_0t-L)^2}, t\in \mathbb{T}_2\cup \mathbb{T}_3. \tag{15b}
\end{equation}

Therefore, the transmission process can be modeled as
\begin{equation*}
y(t)=\left\{
\begin{aligned}
&\sqrt{\frac{h(t)w_s(t)p_0}{d(t)^{\alpha_0}}}x(t)+n(t), t\in \mathbb{T}_1\cup \mathbb{T}_3,  \qquad  \,   (16a)\\
&\sum_{i=1}^2\sqrt{\frac{h_i(t)w^i_s(t)p_0}{d_i(t)^{\alpha_0}}}x_i(t)+n(t), t\in \mathbb{T}_2, \qquad    (16b)\\
\end{aligned}
\right.
\end{equation*}
where $h$ is the channel fading coefficient and $w_s$ is the selected beamforming weight. $\alpha_0$ is the path loss exponent and usually  $\alpha_0\in [2, 5]$. $p_0$ is the average transmit power at HST $H_1$ and $H_2$. $n(t)$ is the additive white Gaussian noise with zero mean and $\sigma_0^2$. Since LOS and large-scale fading is always dominant in HST scenario, it is suggested to assume $h(t)=1$ in the sequel \cite{tt}.

\subsection{Problem Formation}

Let $R_1(t)$ and $R_2(t)$ be the instantaneous information rate that $H_1$ and $H_2$ can achieve at system time $t$, respectively. According to the results of multiple access channel, we know that $R_1(t)$ and $R_2(t)$ satisfy the following inequalities,

\begin{align*}\label{rc}
R_1(t)\le \log_2(1+\frac{w^1_s(t)p_0}{d_1(t)^{\alpha_0}\sigma_0^2}), t\in \mathbb{T}_1\cup \mathbb{T}_2, \tag{17a}\\
R_2(t)\le \log_2(1+\frac{w^2_s(t)p_0}{d_2(t)^{\alpha_0}\sigma_0^2}), t\in \mathbb{T}_2\cup \mathbb{T}_3, \tag{17b}\\
\sum_{i=1}^2R_i(t)\le  \log_2(1+\sum_{i=1}^2\frac{w^i_s(t)p_0}{d_i(t)^{\alpha_0}\sigma_0^2}), t\in \mathbb{T}_2. \tag{17c}\\
\end{align*}

For simplicity, some definitions will be exhibited in advance.

\begin{defn}In the aforementioned discussed scenarios that two trains are served by one common BS, a rate pair ($R_1, R_2$) can be reachable if the following constraints are satisfied.

\begin{equation*}\tag{18}
\left\{
\begin{aligned}
&R_1\le \min \{R_1(t):t\in \mathbb{T}_1\cup \mathbb{T}_2\}\\
&R_2\le \min \{R_2(t):t\in \mathbb{T}_2\cup \mathbb{T}_3\}\\
\end{aligned}
\right.
\end{equation*}

\end{defn}

\begin{defn}In the aforementioned discussed scenarios that two trains are served by one common BS, the achievable rate region $\mathcal{R}$ is the closure of the set of the achievable rate pairs ($R_1, R_2$) with the same average power constraints under arbitrary transmission strategy, that is,

\begin{equation*}\tag{19}
\mathcal{R}=\bigcup\limits_{\{w_1^s(t),w_2^s(t)\}} \{(R_1, R_2)\}.
\end{equation*}

\end{defn}

Consequently, the maximal boundary of $\mathcal{R}$ can be employed to quantify the performance degradation triggered by the encountering process. Because two HSTs have to completely share the wireless channel resource during $t\in \mathbb{T}_2$ and each of them has the common constraints on transmit powers, there exists a tradeoff between $R_1$ and $R_2$. Therefore, it is a multi-objective optimization to maximize the $\mathcal{R}$. One feasible way is adopting the alternative approach iteratively. That is, to approach it by fixing one object first (e.g. fix the data rate $R_2\le R_{max}$), and then maximizing another one object. Both of the objects are alternatively updated. By utilizing the method as in \cite{outage}, we can obtain the maximal achievable rate $R_{max}$ in Eq. (16a) (taking $H_2$ as an example), which is

\begin{equation*}\tag{20}
R_{max}=\log_2(1+\frac{p_0\cdot2L/v_0}{\int_0^{2L/v_0}d_2(t)^{\alpha_0}\sigma_0^2/w^2_s dt}).
\end{equation*}

Based on this result, we define the conditional capacity function $C_{R_2}$.
\begin{defn}When the data rate of $H_2$ is $R_2$, the conditional capacity function $C_{R_2}$ represents the maximal achievable rate of $H_1$.
\end{defn}

According to this definition, the uplink achievable rate region $\mathcal{R}$ can be rewritten as

\begin{equation*}\tag{21}
\mathcal{R}=\{(R_1, R_2)|0\le R_1\le C_{R_2}, 0\le R_2\le R_{max}\}.
\end{equation*}

Consequently, the original multi-objective optimization problem is degraded into a single objective optimization version, which can be rewritten as

\begin{align*}\label{equ:formulation1}
\tag{22}C_{R_2}&=\max_{\{w^1_s(t),w^2_s(t)\}} {R_1}\\
\tag{22a}s.t. \quad &\frac{v_0}{2L}\int_{\mathbb{T}_1\cup \mathbb{T}_2}p_1(t)dt\le p_0,\\
\tag{22b} &\frac{v_0}{2L}\int_{\mathbb{T}_2\cup \mathbb{T}_3}p_2(t)dt\le p_0.\\
\end{align*}

\subsection{Problem Solution}

According to Eq. (12), solving the optimal resource allocation for uplink beamforming in this case is equivalent to finding out the optimal amplitude excitation $f^i_s$ of the selected beam. Thus, the beamforming optimization problem can be rewritten as
\begin{align*}\label{equ:formulation2}
\tag{23}C_{R_2}&=\max_{\{f^1_s(t),f^2_s(t)\}} {R_1}\\
\tag{23a}s.t. \quad &\frac{v_0}{2L}\int_{\mathbb{T}_1\cup \mathbb{T}_2}f^1_s(t)dt\le 1,\\
\tag{23b} &\frac{v_0}{2L}\int_{\mathbb{T}_2\cup \mathbb{T}_3}f^2_s(t)dt\le 1.\\
\end{align*}

Within the effective region $0\le R_2\le R_{max}$, the optimization in Eq. (23) should be considered piecewisely. Intuitively, adaptive resource allocation can enhance the system performance, while the corresponding appropriate arrangement on the order of encoding and decoding following the principle of information theory is very useful in the system design. Otherwise, it may have a negative impact on the rate pair ($R_1, R_2$). That is, for example in superposition coding, the lastly decoded information flow can get a higher energy efficiency. During the encountering process, if only one of the HSTs ($H_1$ or $H_2$) has the information transmission priority, where the priority can stand for the occasion that one of the HSTs has more delay-intolerant information urgent to be transmitted, the maximal achievable rate $R^{'}_{max}$ can be expressed as the following Proposition.

\begin{prop}If $H_2$ has the priority during the encountering process, the maximal achievable rate of $R_1$ under the power constraint in Eq. (23a-23b) is
\begin{equation*}\tag{24}
R^{'}_{max}=\log_2(1+\frac{p_0\cdot2L/v_0}{M\cdot\int_{\mathbb{T}_2}\frac{d_1(t)^{\alpha_0}\sigma_0^2}{w^1_s(t)}dt+\int_{\mathbb{T}_1}\frac{d_1(t)^{\alpha_0}\sigma_0^2}{w^1_s(t)}dt}),
\end{equation*}
where
\begin{equation*}\tag{25}
M=1+\frac{p_0\cdot2L/v_0}{\int_{\mathbb{T}_2}\frac{d_2(t)^{\alpha_0}\sigma_0^2}{w^2_s(t)}dt}.
\end{equation*}
\end{prop}

The Proposition 1 can be simply derived by using the results in \cite{theory}. Likewise, when $H_1$ has the priority during the encountering process, it can be discussed as above.

The rate region when neither of the HSTs has transmission priority will be given in the following Proposition.

\begin{prop}Let $\mathbb{T}_{21}$ and $\mathbb{T}_{22}$ denote the range of $[0, \lambda L/v_0]$ and $[\lambda L/v_0, (2-\eta)L/v_0]$, respectively. If the data rate of $R_2$ is given, the optimal resource allocation of uplink beamforming for $H_1$ and $H_2$ are
\begin{equation*}\tag{26}
f^1_s(t)=\left\{
\begin{aligned}
&\frac{d_1(t)^{\alpha_0}\sigma_0^2}{D^1_s p_0}(2^{C_{R_2}}-1), t\in \mathbb{T}_{1}\cup \mathbb{T}_{22}\\
&\frac{d_1(t)^{\alpha_0}\sigma_0^2}{D^1_s p_0}(2^{C_{R_2}}-1)\cdot 2^{R_2},  t\in \mathbb{T}_{21}\\
\end{aligned}
\right.
\end{equation*}
\begin{equation*}\tag{27}
f^2_s(t)=\left\{
\begin{aligned}
&\frac{d_2(t)^{\alpha_0}\sigma_0^2}{D^2_s p_0}(2^{R_2}-1), t\in \mathbb{T}_{3}\cup \mathbb{T}_{21}\\
&\frac{d_2(t)^{\alpha_0}\sigma_0^2}{D^2_s p_0}(2^{R_2}-1)\cdot 2^{C_{R_2}},  t\in \mathbb{T}_{22}\\
\end{aligned}
\right.
\end{equation*}

The expression of $C_{R_2}$ under power constraint is
\begin{equation*}\tag{28}
C_{R_2}=\log_2(1+\frac{p_0\cdot2L/v_0}{\int_{\mathbb{T}_3\cup \mathbb{T}_{21}}\frac{d_2(t)^{\alpha_0}\sigma_0^2}{w^2_s(t)}dt+M^{'}\cdot\int_{\mathbb{T}_{22}}\frac{d_2(t)^{\alpha_0}\sigma_0^2}{w^2_s(t)}dt}),
\end{equation*}
where
\begin{equation*}\tag{29}
M=1+\frac{p_0\cdot2L/v_0}{\int_{\mathbb{T}_{22}}\frac{d_1(t)^{\alpha_0}\sigma_0^2}{w^1_s(t)}dt},
\end{equation*}
and the value of $\lambda$ can be recursively calculated by the average power constraint in Eq. (23a-23b).
\end{prop}

The proof of Proposition 2 will be given in appendix A. {\hfill $\blacksquare$}

In fact, to achieve the maximum data rate of $R_1$, the decoding strategy should be appropriately designed. Namely, if the resource allocation beamforming amplitude excitation for $H_2$ is high enough to meet the data rate $R_2$, $H_2$ needs to give up the decoding priority during the encountering period. Otherwise, the decoding priority of $H_2$ is necessary in order to meet the data rate requirement. This conclusion in Proposition 2 can be extended to the occasion when $R_1$ is given.

To conclude it, the location-aided adaptive optimal resource allocation of uplink beamforming scheme to maximize the achievable rate region can be attained by using the results of Proposition 1 and 2, which can substantially alleviate the performance degradation introduced by encountering.
%% ÊýÖµ½á¹û£º
\section{Numerical Results}\label{Sec:NumRes}
In this section, numerical results are presented. Assume that $d_0=50$ m, $h_0=20$ m, $v_0=360$ km/h, $L=800$ m, $\alpha_0=3$, $f_c = 2.4$ GHz, $\lambda = \frac{1}{f_c}$, $d = \frac{\lambda}{2}$ and $N=128$.

\begin{figure}[b]
\centering
\includegraphics[width=0.5\textwidth]{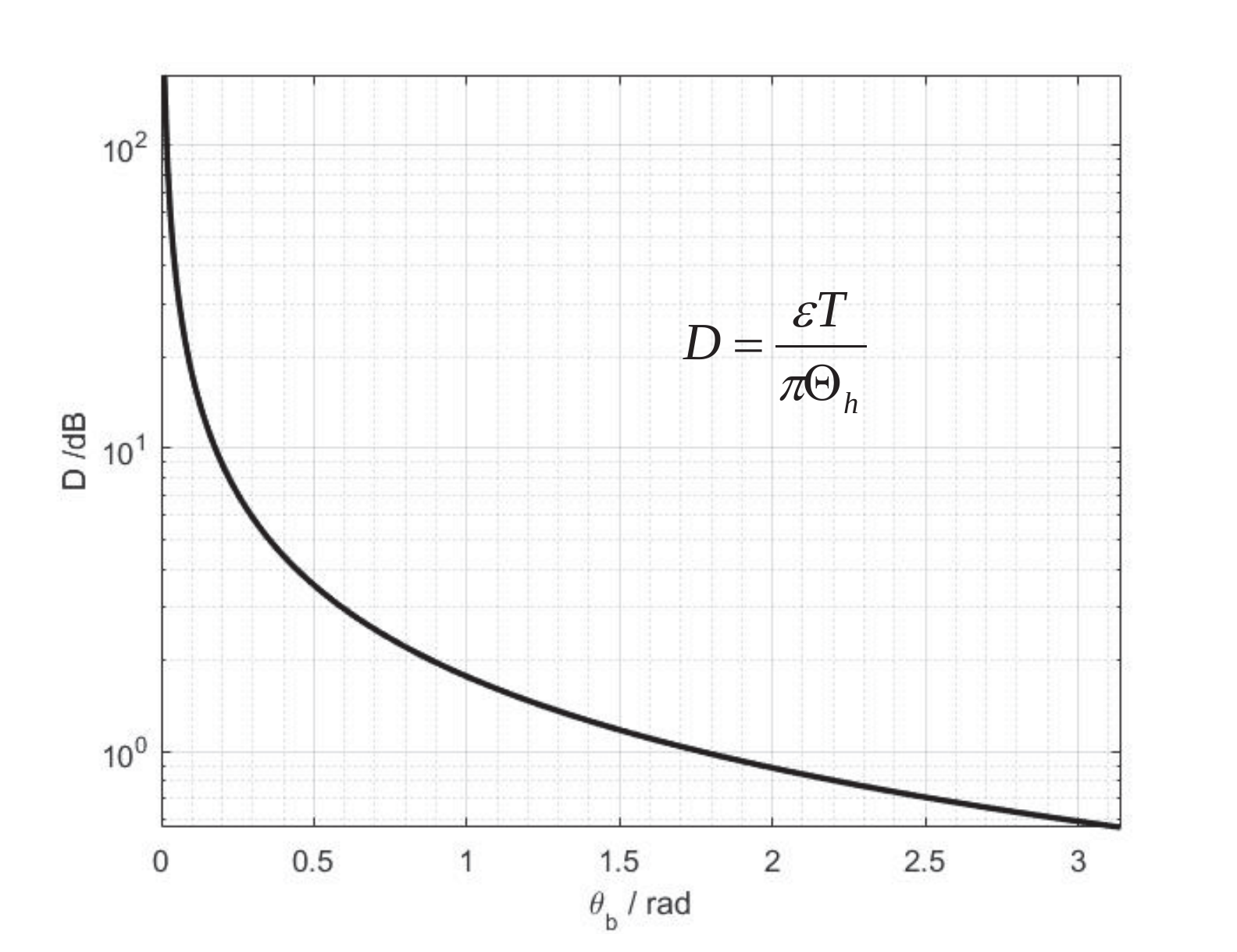}
\caption{The tradeoff between beam directivity and beamwidth.} \label{Fig:Tradeoff}
\end{figure}

The tradeoff between $\Theta_h$ and $D$ is shown in Fig. \ref{Fig:Tradeoff}, where for a given beamwidth restricted by the positioning error, the corresponding beam directivity is shown. It can be observed that the beam directivity $D$ decreases with respect to beamwidth $\Theta_h$, where $\Theta_h$ varies from $0$ to $\pi$. It also indicates that the beam directivity increases with total beam number.

\begin{figure}[t]
\centering
\includegraphics[width=0.5\textwidth]{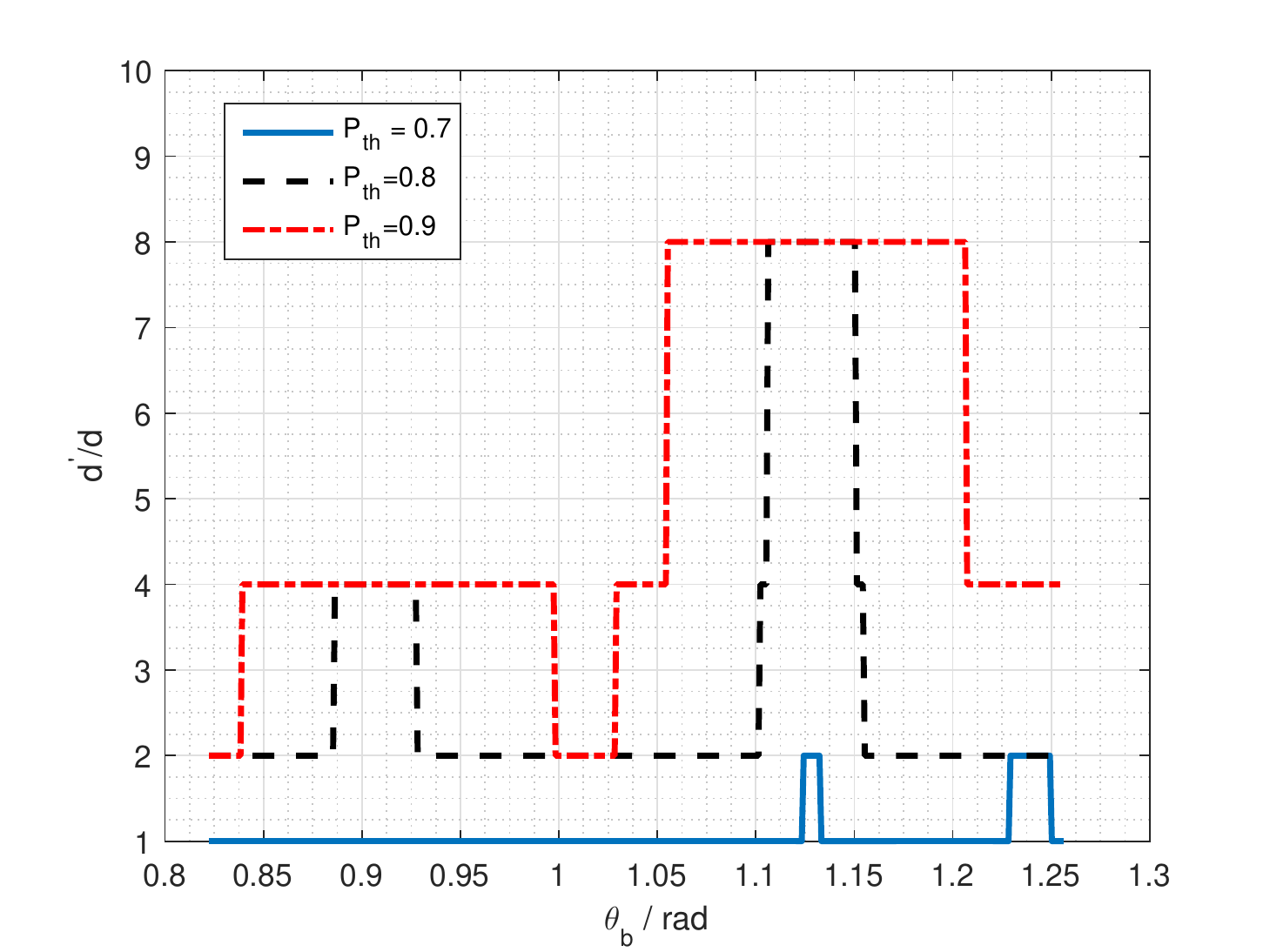}
\caption{The variation of $d$ versus base station $\theta_b$ when the directivity is constant and $P_{th}$ = 0.7, 0.8 and 0.9.} \label{Fig:bVersusThetab}
\end{figure}

As shown in Fig. \ref{Fig:bVersusThetab}, when $P_{th}$ = 0.7, 0.8 and 0.9, the variation of directivity versus train location $\theta_b$ has been depicted. It can be observed that the directivity varies with respect to $\theta_b$, but there exists no monotonicity. When the BS is near to the edge of BS coverage, to guarantee the common beamforming rate, the beam number tends to be relatively large. However, when the BS is near the center of one beam, the beam number tends to be relatively small. It reflects the adaptation of directivity-beamwidth tradeoff in our optimization to guarantee diverse thresholds.

\begin{figure}[b]
\centering
\includegraphics[width=0.5\textwidth]{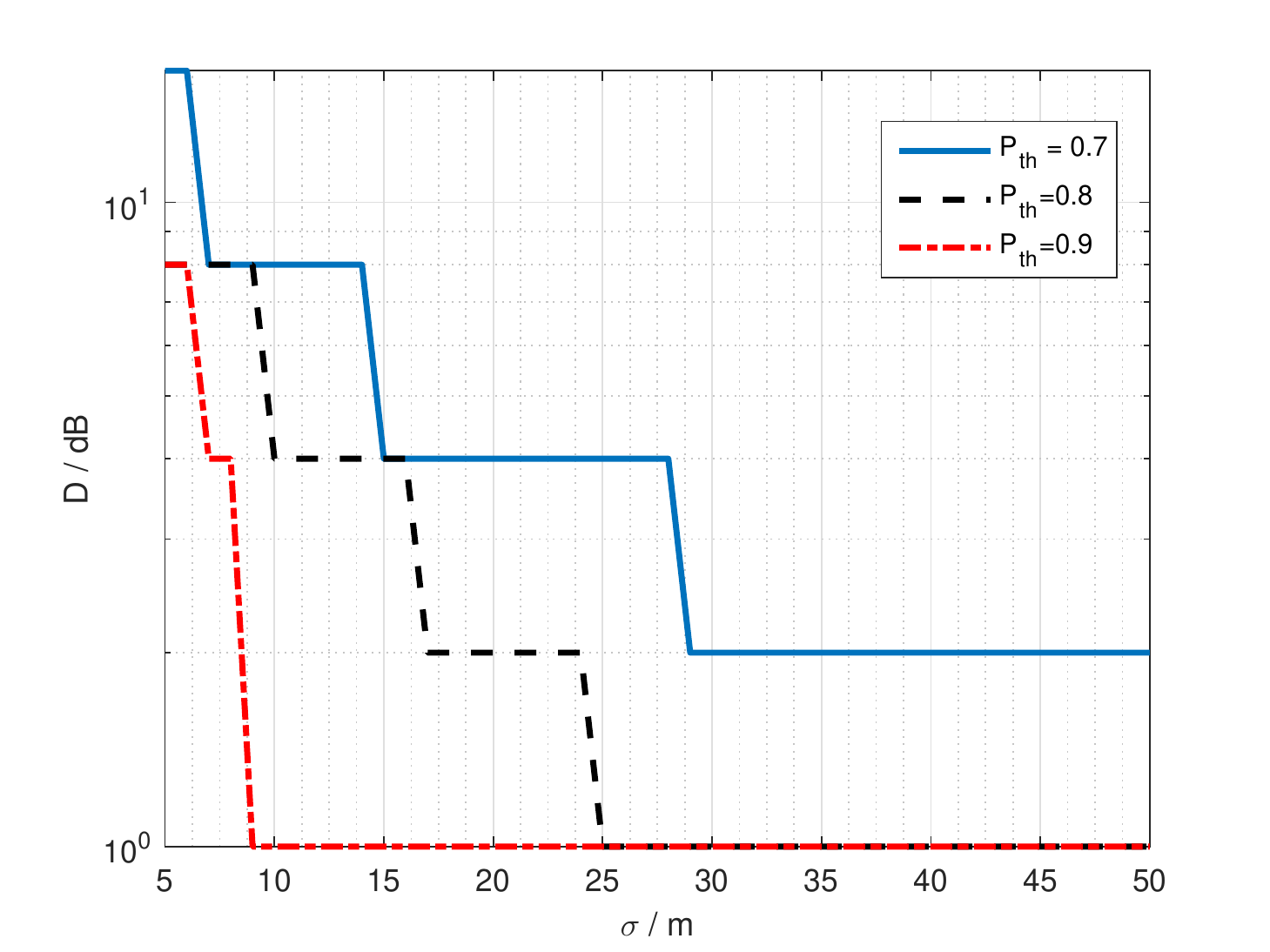}
\caption{The variation of directivity versus positioning error variance $\sigma$ when $P_{th}$ = 0.7, 0.8 and 0.9; $\theta_b=\frac{\pi}{4}$.} \label{Fig:DversusSigma}
\end{figure}

In addition, Fig. \ref{Fig:bVersusThetab} shows that when the directivity is fixed, the variation of $\frac{d}{d^{'}}$ is exactly the inverse of $\frac{N^{'}}{N}$, where $d^{'}$ represents the altered antenna spacing. The numerical results agree with that expressed in $\bold{Lemma}$ 1. It also indicates that to guarantee a better robustness, the antenna spacing needs to be designed as large as possible, which reduces the implementation complexity owing to the large space on the top of train.

The variation of directivity versus positioning error variance $\sigma$, when $P_{th}$ = 0.7, 0.8 and 0.9 and $\theta_b=\frac{\pi}{4}$, are shown in Fig. \ref{Fig:DversusSigma}. The directivity decreases linearly with $\sigma$ but increases with $P_{th}$ because larger $\sigma$ or $P_{th}$ means smaller beam number will be adopted so that the switching times among the beams will be decreased. That is, the switching drop probability will be reduced.

\begin{figure}[t]
\centering
\includegraphics[width=0.5\textwidth]{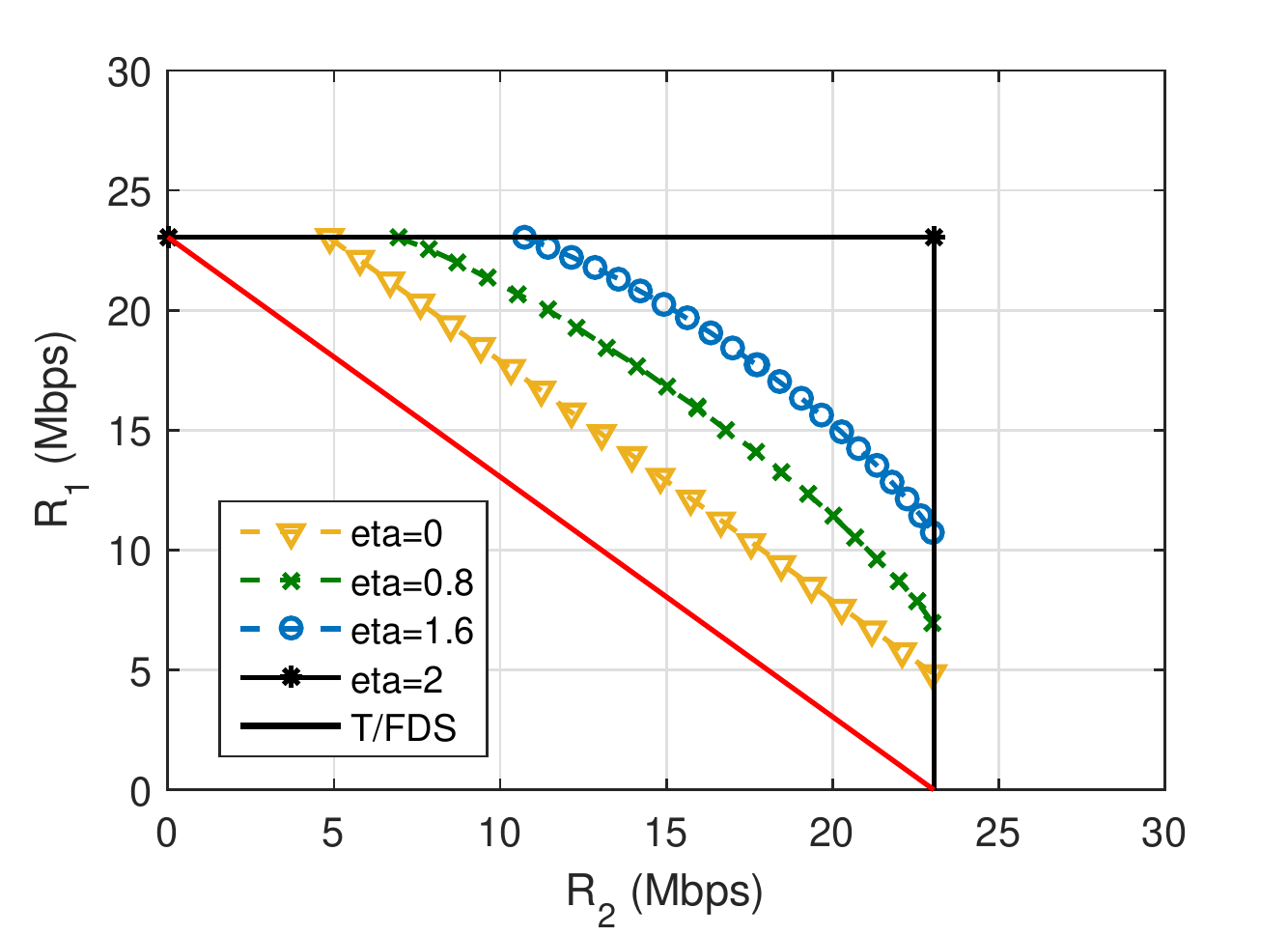}
\caption{The achievable rate region of uplink beamforming for two HSTs encountering scenario, in which $\eta=0, \eta=0.8, \eta=1.6$ and $\eta=2$.} \label{Fig:region}
\end{figure}

\begin{figure}[b]
\centering
\includegraphics[width=0.5\textwidth]{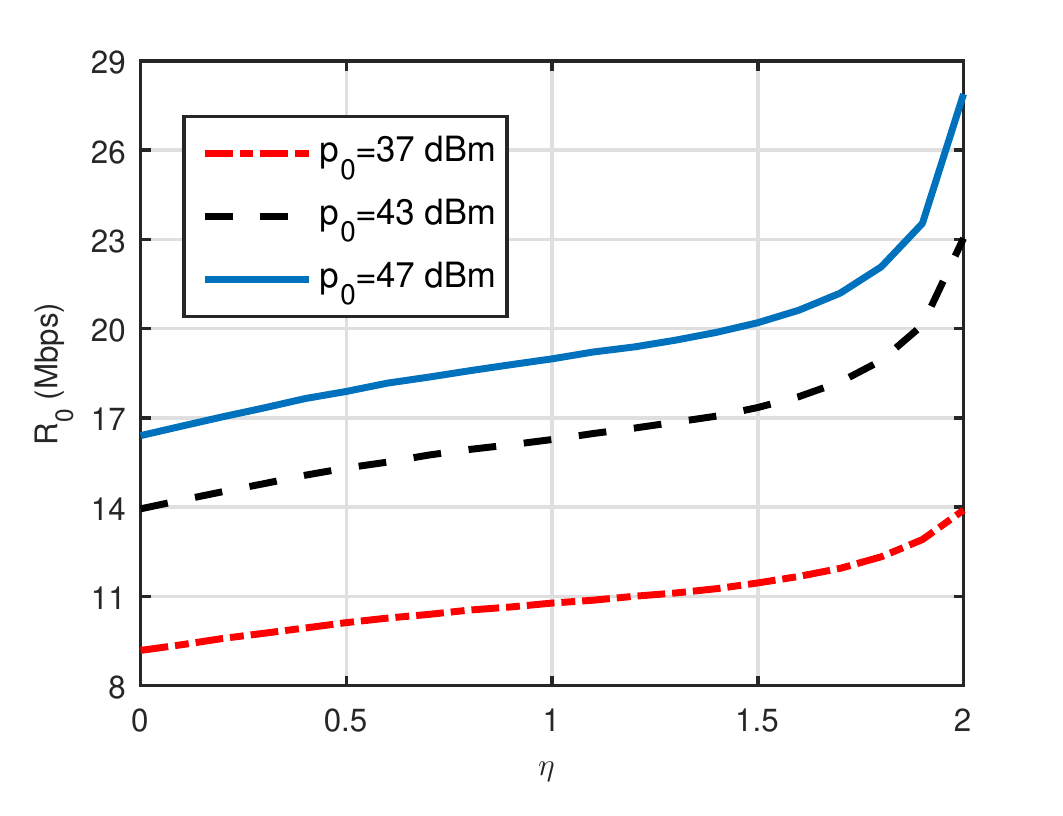}
\caption{The relation between $\eta$ and the data rate of a symmetric system, where the transmit power constraints are $p_0=$37 dBm, $p_0=$43 dBm, $p_0=$47 dBm.} \label{Fig:eta}
\end{figure}

The achievable rate region in the two HSTs encountering scenario is illustrated in Fig. \ref{Fig:region}, which shows four cases: $\eta=0, \eta=0.8, \eta=1.6$ and $\eta=2$. When $\eta=0$, it stands for the worst case that the encountering time period of the two HSTs is the longest. When $\eta=2$, it represents the two HSTs are served by different BSs, which achieves the largest achievable rate region. To exhibit the performance improvement, the achievable rate region of time/frequency division scheme (T/FDS) is presented (red line). It is obvious that compared with the worst case $\eta=0$, due to the adoption of adaptive resource allocation of uplink beamforming and optimal decoding, the performance improvement is quite large.

In Fig. \ref{Fig:eta}, a symmetric system is considered to study the performance degradation over encountering time, where both of the HSTs demand for the same data rate ($R_1=R_2=R_0$) in the encountering process. Three different transmit power constraints ($p_0=$37 dBm, 43 dBm, 47 dBm) are illustrated with the same resource allocation scheme for uplink beamforming. According to the simulation result, $R_0$ is a monotonous function of $\eta$. Therefore, the $\eta$ value can be treated as an impact factor on the wireless data rate during the encountering period, where the bigger the $\eta$ value, the less impact. In fact, the $\eta$ value can be predicted according to the HST location and velocity, which indicates that the potential optimal transmission strategy can be determined according to the amount and type of the real-time wireless transmission demands.

\section{Conclusions}

In this paper, we first analyzed the beamforming design principles for massive MIMO system based on location information and then presented a low-complexity beamforming implementation scheme in high mobility scenario. Different from conventional beamforming schemes, our design needs neither acquiring UCCM and DCCM nor ED them, which not only substantially reduces the system complexity and on-line computational complexity, but also possesses a favourable robustness to the CSI estimations. Therefore, the proposed beamforming scheme can benefit the design of wireless communication system for HST. It is noted that the location information plays a paramount role in our scheme, where the tradeoff between beamwidth and directivity in this scenario and how to maximize direcitivity under diverse positioning accuracies to guarantee efficient transmission are crucial, especially in engineering design of HST wireless communication systems. Finally, aiming to solve the system performance deterioration in future high traffic density HST network, an explicit closed-form expression of optimal location-aided resource allocation of uplink beamforming scheme was presented when two HSTs need to be served by one BS. By utilizing the HST location information, we can adjust the beamforming scheme according to the estimated encountering time of two HSTs, which can alleviate the performance degradation during their encountering period according to different transmission priorities.

\section*{Acknowledgement}

This work was supported by State Key Development Program of Basic Research of China No. 2012CB316100(2) and National Natural Science Foundation of China (NSFC) No. 61321061.
% You must have at least 2 lines in the paragraph with the drop letter
% (should never be an issue)

\begin{appendices}

\section{}\label{Appen:pro1}

In the time period $\mathbb{T}_{21}$, $H_2$ has the decoding priority and In the time period $\mathbb{T}_{22}$, $H_1$ has the decoding priority. During the time period $t\in \mathbb{T}_{21}\cup \mathbb{T}_{1}$ for $H_2$ and $t\in \mathbb{T}_{3}\cup \mathbb{T}_{22}$ for $H_1$, it is similar to a single HST scenario, where the optimal resource allocation of beamforming scheme is similar as in \cite{outage}. Therefore, for $H_1$ ($t\in \mathbb{T}_{21}\cup \mathbb{T}_{1}$), the power allocation coefficient is

\begin{equation*}
f^1_s=\frac{d_1(t)\sigma_0^2}{D^1_s p_0}(2^{C_{R_{2}}}-1).
\end{equation*}

Similarly, the power allocation coefficient for $H_2$ when $t\in \mathbb{T}_{21}\cup \mathbb{T}_{1}$ is
\begin{equation*}
f^2_s=\frac{d_2(t)\sigma_0^2}{D^2_s p_0}(2^{R_{2}}-1).
\end{equation*}

In the time period $\mathbb{T}_{21}$ for $H_1$, since the $H_2$ possesses a high decoding priority, the transmission from $H_2$ is treated as noise in the decoding process. Thus, the optimal resource allocation of beamforming scheme $f^1_s$ for $H_1$ can be achieved with fixed data rate $R_2$ as in \cite{david}, which is similar to our previous work in \cite{tt} and therefore, can be expressed as

\begin{equation*}
f^1_s=\frac{d_1(t)^{\alpha_0}\sigma_0^2}{D^1_s p_0}(2^{C_{R_2}}-1)\cdot 2^{R_2}.
\end{equation*}

It is the same when $H_1$ has the decoding priority in time period $\mathbb{T}_{22}$, where the optimal resource allocation of beamforming scheme $f^2_s$ is

\begin{equation*}
f^2_s=\frac{d_2(t)^{\alpha_0}\sigma_0^2}{D^2_s p_0}(2^{R_2}-1)\cdot 2^{C_{R_2}}.
\end{equation*}

This completes the proof of Proposition 2.
\end{appendices}


\begin{thebibliography}{1}

\bibitem{5G}
J. G. Andrews \emph{et al.}, \textquotedblleft What will 5G be?" \emph{IEEE J. Sel. Areas Commun.}, vol. 32, no. 6, pp. 1065--1082, Jun. 2014.

\bibitem{seamless}
O. B. Karimi, J. C. Liu, and C. G. Wang, \textquotedblleft Seamless wireless connectivity for multimedia services in high speed trains," \emph{IEEE J. Sel. Areas Commun.}, vol. 30, no. 4, pp. 729--739, May. 2012.

\bibitem{5G1}
F. Boccardi, R. W. Heath, A. Lozano, T. L. Marzetta, and P. Popovski, \textquotedblleft Five disruptive technology directions for 5G," \emph{IEEE Commun. Mag.}, vol. 52, no. 2, pp. 74--80, Feb. 2014.

\bibitem{beamform}
X. Gao, L. Dai, C. Yuen, and Z. Wang, \textquotedblleft Turbo-Like beam-forming Based on Tabu Search Algorithm for Millimeter-Wave Massive MIMO Systems," \emph{IEEE Trans. Veh. Technol.}, vol. 65, no. 7, pp. 5731--5737, Jul. 2016.

\bibitem{work2}
J. Li, Z. Zhang, C. Qi, C. Wang, and C. Zhong, \textquotedblleft Gradual beam-forming and soft handover in high mobility cellular communication networks," \emph{in Proc. IEEE Globecom Workshops}, Atlanta, USA, Dec. 2013, pp. 1320--1325.

\bibitem{work3}
M. Cheng, X. Fang, and W. Luo, \textquotedblleft beam-forming and positioning-assisted handover scheme for long-term evolution system in high-speed railway," \emph{IET Commu.}, vol. 6, no. 15, pp. 2335--2340, Oct. 2012.

\bibitem{work4}
X. Chen, S. Liu, and P. Fan, \textquotedblleft Position-based diversity and multiplexing analysis for high speed railway communications," \emph{in Proc. IEEE Workshop on High Mobility Wireless Commun. (HMWC 2015)}, Xi'an, China, Oct. 2015, pp. 41--45.

\bibitem{mobility}
F. B. Tesema, A, Awada, I. Viering, M. Simsek, and G.P. Fettweis, \textquotedblleft Mobility modeling and performance evaluation of multi-connectivity in 5G intra-frequency networks," \emph{in Proc. IEEE Globecom Workshops}, San Diego, USA, Dec. 2015, pp. 1--6.

\bibitem{mobility1}
H. Song, X. Fang, and L. Yan , \textquotedblleft Handover Scheme for 5G C/U Plane Split Heterogeneous Network in High-Speed Railway," \emph{IEEE Trans. Veh. Technol.}, vol. 63, no. 9, pp. 4633--4646, Nov. 2014.

\bibitem{demand}
L. Liu, C. Tao, et al, \textquotedblleft Position-based modeling for wireless channel on high-speed railway under a viaduct at 2.35 GHz," \emph{IEEE J. Sel. Areas Commun.}, Vol. 30, No. 4, pp. 834--845, Apr. 2012.

\bibitem{work}
X. Chen, S. Liu, J. Lu, P. Fan and K. B. Letaief, \textquotedblleft Smart channel sounder for 5G IoT: from wireless big data to active communication," \emph{IEEE ACCESS}, Vol. 4, pp. 8888--8899, Nov. 2016.

\bibitem{work1}
J. Lu, X. Chen, S. Liu, and P. Fan, \textquotedblleft Location-aware low complexity ICI reduction in OFDM downlinks for high-speed railway communication systems with distributed antennas," \emph{in Proc. IEEE VTC2016-Spring}, Nanjing, China, May. 2016, pp. 1--5.

\bibitem{abf}
L. Griffiths, C. W. Jim, \textquotedblleft An alternative approach to linearly constrained adaptive beam-forming," \emph{IEEE Trans. Antennas Propag.}, vol. 30, no. 1, pp. 27--34, Jul. 1982.

\bibitem{obf}
Z. Lei, F. P. S. Chin and Y. Liang, \textquotedblleft Orthogonal switched beams for downlink diversity transmission," \emph{IEEE Trans. Antennas Propag.}, vol. 53, no. 7, pp. 2169--2177, Jul. 2005.

\bibitem{gain}
M. Cheng, X. Fang, L. Yan, \textquotedblleft Beam-forming and Alamouti STBC combined downlink transmission schemes in communication systems for high-speed railway," \emph{in Proc. International Conference on Wireless Communications and Signal Processing (WCSP)}, Hangzhou, pp. 1-6, 2013.

%\bibitem{work5}
%X. Chen, J. Lu, S. Liu and P. Fan, \textquotedblleft Location-Aided Umbrella-Shaped Massive MIMO beam-forming Scheme with Transmit Diversity for High Speed Railway Communications," \emph{in Proc. IEEE VTC2016-Spring}, Nanjing, China, May. 2016, pp. 1-5.

\bibitem{antenna}
C. A. Balanis, \emph{Antenna Theory: Analysis and Design}. New Jersey, USA: John Wiley \& Sons, 2005.

\bibitem{directional}
V. Raghavan, S. Subramanian, J. Cezanne and A. Sampath, \textquotedblleft Directional beamforming for millimeter-wave MIMO systems," \emph{in Proc. IEEE Globecom}, San Diego, CA, USA, Dec. 2015, pp. 1--7.

\bibitem{tt}
T. Li, Z. Chen, P. Fan and K. Ben Letaief, \textquotedblleft Position-based power allocation for uplink HSRs wireless communication when two trains encounter," \emph{in Proc. IEEE Globecom}, Washington, DC, USA, Dec. 2016, pp. 1--6.

\bibitem{outage}
A. Goldsmith and P. Varaiya, \textquotedblleft Capacity of fading channel with channel side information," \emph{IEEE Trans. Inf. Theory}, vol. 43, no. 6, pp. 1986--1992, Nov. 1997.

\bibitem{theory}
T. Cover and J. A. Thomas, \textquotedblleft Element of infomation theory," 2nd ed. Wiley Interscience, 2006.

\bibitem{david}
S. V. Hanly and D. N. C. Tse, \textquotedblleft Multiaccess fading channels¨Cpart II: delay-limited capacities," \emph{IEEE Trans. Inf. Theory}, vol. 44, no. 7, pp. 2816--2831, Nov. 1998.

\end{thebibliography}
\end{document}